\documentclass{article}
\usepackage[dvips]{graphicx}

\input{tcilatex}

\begin{document}

\title{The Instabilities of Bianchi Type IX Einstein Static Universes}
\author{John D Barrow and Kei Yamamoto \\
%EndAName
DAMTP, Centre for Mathematical Sciences,\\
University of Cambridge,\\
Wilberforce Rd., Cambridge CB3 0WA\\
United Kingdom}
\maketitle

\begin{abstract}
We investigate the stability of the Einstein static universe as a non-LRS
Bianchi type IX solution of the Einstein equations in the presence of both
non-tilted and tilted fluids. We find that the static universe is unstable
to homogeneous perturbations of Bianchi type IX to the future and the past. 
\newline

\end{abstract}

%\section{}
%\subsection{}

\section{Introduction}

The stability of the Einstein static universe is a problem with a history as
old as relativistic cosmology. The static universe was the first
cosmological solution of general relativity to be found, by Einstein \cite%
{ein}, and the study of its stability, implicitly by Lema\^{\i}tre \cite%
{lem, lem2}, and then explicitly by Eddington \cite{edd}, played an
important role in establishing the concept of the expanding universe. These
authors only showed that the Einstein static solution was unstable to
spatially homogeneous and isotropic perturbations with a limited range of
perfect fluid equations of state. Nevertheless, after 1930, it was widely
believed that the Einstein static universe was always unstable and would
evolve into a state of universal contraction or expansion if perturbed \cite%
{bonn}. In 1967, Harrison \cite{harr} showed that all the perturbations
which render the dust-filled Einstein static universe unstable are in fact
oscillatory if the dust is replaced by black body radiation, and also showed
that the Einstein static universe is neutrally stable against small
inhomogeneous perturbations for a range of perfect fluids. Similar results
were found for homogeneous and isotropic perturbations of all amplitudes by
Gibbons \cite{gib1, gib2} who also provided a thermodynamic perspective and
showed that if $dp/d\rho >1/5$, where $p$ is the pressure and $\rho $ is the
density of the contents of the universe, then the static universe is stable.
The physical reason for this surprising conclusion is that the Einstein
static space is compact and when the sound speed exceeds $1/\sqrt{5}$ the
Jeans length exceeds the size of the universe and the instability growth is
quenched by the pressure. These studies are quite limited and were extended
by Barrow, Ellis, Maartens and Tsagas \cite{BEMT} to consider the whole
spectrum of inhomogeneous scalar, vector and tensor perturbations to the
static universe, as well as the presence of a self-interacting scalar field.
They found that the static universe is neutrally stable against all
inhomogeneous adiabatic scalar perturbations if $dp/d\rho >1/5$ and
confirmed that the dust ($p=0$) model is always unstable. Any initial vector
perturbations remain frozen in and so there is neutral stability against
vortical perturbations for all equations of state at linear order. Likewise,
there is neutral stability with respect to transverse-traceless tensor
perturbations for all equations of state. Two caveats are to be noted. Since
the Einstein static universe has compact space sections and many Killing
vectors, any perturbation analysis is susceptible to the problems of
linearisation instability. In effect, Einstein static is a conical point in
the space of all solutions to the Einstein equations and special
higher-order constraints must be checked to ensure that the leading terms in
any perturbative series expansion around it are the leading-order terms in
series that converge to true solutions of the Einstein equations, see for
example refs \cite{brill, BTip, BEMT, unr}. Also, as an aside, we note that
if 'ghost' fields with negative densities ($\rho <0$) are permitted then
Einstein's static universe is a stable solution of the field equations and
exact isotropic and homogeneous cosmological solutions can be found that
exhibit stable oscillations of arbitrarily small amplitude around the static
state \cite{BT} -- in effect these are oscillating universes with non-zero
minimum and maximum radii.

Einstein static universes are also particular solutions in a wide range of
extensions of general relativity. Existence conditions are known, for
example, for gravitation theories in which the lagrangian is an arbitrary
function of the scalar, Ricci, and Riemann \ curvature invariants \cite{BO,
BC}. There have been many studies of the Einstein static stability in other
gravity theories \cite{zhang, par, boem, boem2, boem3, tav, ss, dun, bruni},
and from a thermodynamic point of view \cite{vol}. There are also possible
cosmological initial states that rely upon stabilities of the static
universe, for example that of the emergent universe scenario \cite{Emer},
which requires the Einstein static universe to be stable to the infinite
past and expand like an Eddington-Lema\^{\i}tre universe to the future.
Eddington favoured a cosmological scenario of this sort, with a infinite
geometrical past but only a finite thermodynamic past when the universe was
significantly out of equilibrium.

These earlier studies of the stability of Einstein static universes have
examined the behaviour of small and large amplitude homogeneous and
isotropic (conformal) perturbations, and small amplitude inhomogeneous
perturbations. However, in our earlier study of the stability properties in
general relativity, \cite{BEMT}, we also pointed out that a key ingredient
in understanding the stability of Einstein static is to understand its
response to long-wavelength homogeneous gravitational waves of Bianchi type
IX. We showed that there was instability with respect to the mode that
controlled the volume expansion in the non-tilted perfect fluid case. In
this paper we will provide a full analysis of the stability of the Einstein
universe with respect to the non-LRS Bianchi type IX degrees of freedom for
tilted and non-tilted perfect fluids. This problem reduces to studying the
stability properties of the special Einstein static universe solution of the
Bianchi type IX ('Mixmaster') Einstein equations. This anisotropic metric
describes the most general spatially homogeneous closed universe with
Einstein static as a special case\footnote{%
Note that Einstein static is not a special case of the Kantowski-Sachs
closed anisotropic universes, which have $S^{1}\times S^{2}$space sections.}%
. We will show that the static universe is unstable to the past and to the
future and we expect the solution to be unstable in other gravity theories
as well if Mixmaster exact solutions exist. In particular, cosmological
scenarios with a past-eternal 'initial' Einstein static are unstable in
general relativity.

\section{Non-tilted Bianchi Type IX Universes}

We will proceed by showing that the Einstein static universe is a particular
exact Bianchi type IX solution of the Einstein equations for spatially
homogeneous cosmologies and then determine its stability. This will show any
effects of the homogeneous gravitational degrees of freedom present in type
IX universes. This metric possesses anisotropic spatial curvature and is the
only member of the Bianchi classification of homogeneous spaces to contain
the closed Friedmann universes as a special case. We will examine the
situation with a non-tilted perfect fluid source first.

We follow Ellis and MacCallum \cite{mac} to write down the equations in a
group invariant orthonormal frame. The geometry of Bianchi type IX
space-time is characterised by the kinematic quantities of the homogeneous
spatial hypersurface: bulk expansion rate $H$, trace-free shear tensor $%
\sigma _{\alpha \beta }$, an auxiliary three-vector $\Omega _{\alpha }$ that
measures the rotation of the frame with respect to Fermi-propagated one and
a symmetric three-tensor $n_{\alpha \beta }$ which determine the internal
geometry of the spatial hypersurface. The spatial scalar curvature and
trace-free part of Ricci tensor are given by 
\begin{eqnarray*}
^3 R &=& -n_{\mu \nu }n^{\mu \nu } + \frac{1}{2}(n_{\mu }^{\ \mu })^2 \\
^3 S_{\alpha \beta } &=& 2n_{\alpha \mu }n^{\mu }_{\ \beta } - n_{\mu }^{\
\mu } n_{\alpha \beta } -\frac{1}{3}(2n_{\mu \nu }n^{\mu \nu } - (n_{\mu
}^{\ \mu })^2 )\delta _{\alpha \beta } .
\end{eqnarray*}
When the fluid is non-tilted, we can diagonalise $\sigma _{\alpha \beta }$
and $n_{\alpha \beta }$ simultaneously, which also implies $\Omega _{\alpha
}=0$. We assume the fluid satisfies 
\begin{equation}
p = (\gamma -1)\rho  \label{eq:eos}
\end{equation}
with a constant $\gamma $. The cosmological constant $\Lambda $ can be
regarded as another fluid with $\rho _{\Lambda } = -p_{\Lambda } = \Lambda $%
. Using the notations in \cite{HUW}, there are eight variables $\{ H, \sigma
_{\pm }, n_{1,2,3}, \rho , \rho _{\Lambda } \}$. For Bianchi  type IX, $%
n_{1,2,3}$ are all non-zero and of a same signature, here taken to be
positive.

To obtain dimensionless equations, let us define a new variable 
\[
D\equiv \sqrt{H^{2}+\frac{1}{4}(n_{1}n_{2}n_{3})^{\frac{2}{3}}}, 
\]%
which we will use to normalise the system of equations for Bianchi IX,
following Hewitt, Uggla and Wainwright \cite{HUW}. The advantage of using
this variable is that the normalisation with respect to $D$ is well-defined
everywhere in the type IX state space since $D \neq 0$ as long as $n_{1,2,3}
\neq 0$. Particularly, it is non-zero for Einstein static universe where $H=0
$.

We introduce dimensionless variables as follows: \emph{\ } 
\begin{equation}
\tilde{H}\equiv \frac{H}{D},\ \ \tilde{\Sigma }_{\pm }\equiv \frac{\sigma
_{\pm }}{D},\ \ \tilde{N}_{i}\equiv \frac{n_{i}}{D},\ \ \tilde{\Omega }%
_{m}\equiv \frac{\rho }{3D^{2}},\ \ \tilde{\Omega }_{\Lambda }\equiv \frac{%
\Lambda }{3D^{2}} .
\end{equation}%
We also introduce a new time variable $\tau $ by defining

\bigskip 
\[
\frac{dt}{d\tilde{\tau }}=\frac{1}{D}>0 
\]%
and denote $d/d\tau $ by $^{\prime }$.

We may use the normalised equations given in \cite{HUW} with a slight
modification to accommodate two fluids. First of all, the evolution of $D$
decouples as 
\begin{equation}
D^{\prime }=-(1+\tilde{q})\tilde{H}D
\end{equation}%
where $\tilde{q}$ is algebraically determined by the normalised variables
through Raychaudhuri equation 
\begin{equation}
\tilde{q}=2(\tilde{\Sigma}_{+}^{2}+\tilde{\Sigma}_{-}^{2})+\frac{1}{2}%
(3\gamma -2)\tilde{\Omega}_{M}-\tilde{\Omega}_{\Lambda }.
\end{equation}%
The rest of the evolution equations read 
\begin{eqnarray}
\tilde{H}^{\prime } &=&-(1-\tilde{H}^{2})\tilde{q}, \\
\tilde{\Sigma}_{\pm }^{\prime } &=&-(2-q)\tilde{H}\tilde{\Sigma}_{\pm }-%
\tilde{S}_{\pm }, \\
\tilde{N}_{1}^{\prime } &=&(\tilde{H}\tilde{q}-4\Sigma _{+})\tilde{N}_{1}, \\
\tilde{N}_{2}^{\prime } &=&(\tilde{H}\tilde{q}+2\tilde{\Sigma}_{+}+2\sqrt{3}%
\tilde{\Sigma}_{-})\tilde{N}_{2}, \\
\tilde{N}_{3}^{\prime } &=&(\tilde{H}\tilde{q}+2\tilde{\Sigma}_{+}-2\sqrt{3}%
\tilde{\Sigma}_{-})\tilde{N}_{3}, \\
\tilde{\Omega}_{M}^{\prime } &=&(2\tilde{q}-3\gamma +2)\tilde{H}\tilde{\Omega%
}_{M}, \\
\tilde{\Omega}_{\Lambda }^{\prime } &=&2(\tilde{q}+1)\tilde{H}\tilde{\Omega}%
_{\Lambda }.
\end{eqnarray}%
The components of spatial Ricci tensor are given by 
\begin{eqnarray*}
\tilde{S}_{+} &=&\frac{1}{6}\left[ (\tilde{N}_{2}-\tilde{N}_{3})^{2}-\tilde{N%
}_{1}(2\tilde{N}_{1}-\tilde{N}_{2}-\tilde{N}_{3})\right] , \\
\tilde{S}_{-} &=&\frac{1}{2\sqrt{3}}\left[ (\tilde{N}_{2}-\tilde{N}_{3})(%
\tilde{N}_{1}-\tilde{N}_{2}-\tilde{N}_{3})\right] .
\end{eqnarray*}%
There are two constraint equations 
\begin{eqnarray}
1 &=&\tilde{\Sigma}_{+}^{2}+\tilde{\Sigma}_{-}^{2}+\tilde{V}+\tilde{\Omega}%
_{M}+\tilde{\Omega}_{\Lambda },  \label{eq:const2} \\
1 &=&\tilde{H}^{2}+\frac{1}{4}\left( \tilde{N}_{1}\tilde{N}_{2}\tilde{N}%
_{3}\right) ^{\frac{2}{3}},  \label{eq:const1}
\end{eqnarray}%
where 
\begin{equation}
\tilde{V}=\frac{1}{12}\left[ \tilde{N}_{1}^{2}+\tilde{N}_{2}^{2}+\tilde{N}%
_{3}^{2}-2\tilde{N}_{1}\tilde{N}_{2}-2\tilde{N}_{2}\tilde{N}_{3}-2\tilde{N}%
_{3}\tilde{N}_{1}+3\left( \tilde{N}_{1}\tilde{N}_{2}\tilde{N}_{3}\right) ^{%
\frac{2}{3}}\right] .
\end{equation}

\section{The Stability of Einstein Static Universes}

The only equilibrium point in the interior of type IX invariant set
corresponds to the Einstein static solution, given by 
\begin{equation}
\tilde{H}=\tilde{\Sigma }_{\pm } =0, \ \ \ \ \ \tilde{N}_{1,2,3} = 2, \ \ \
\ \ \tilde{\Omega }_{M} = \frac{2}{3\gamma }, \ \ \ \ \ \tilde{\Omega }%
_{\Lambda } = 1-\frac{2}{3\gamma }.
\end{equation}
We are interested in linear stability around it.

\textbf{Linearisation}:

We denote a small deviation of a quantity $x$ from its value at the Einstein
static solution, $x_{0}$, by $\delta x=x-x_{0}$. From (\ref{eq:const1}), we
have 
\begin{equation}
\delta \tilde{N}_{1}+\delta \tilde{N}_{2}+\delta \tilde{N}_{3}=0,
\label{eq:deltaN}
\end{equation}%
and, with (\ref{eq:const2}), this means that 
\begin{equation}
\delta \tilde{V}=0,\ \ \ \ \ \delta \tilde{\Omega}_{M}=-\delta \tilde{\Omega}%
_{\Lambda }.
\end{equation}%
From the definitions of $\tilde{q}$ and $\tilde{S}_{\pm }$, we have 
\begin{eqnarray}
\delta \tilde{q} &=&\frac{1}{2}(3\gamma -2)\delta \tilde{\Omega}_{M}-\delta 
\tilde{\Omega}_{\Lambda } \\
\delta \tilde{S}_{+} &=&-\frac{1}{3}(2\delta \tilde{N}_{1}-\delta \tilde{N}%
_{2}-\delta \tilde{N}_{3})=\delta \tilde{N}_{2}+\delta \tilde{N}_{3} \\
\delta \tilde{S}_{-} &=&-\frac{1}{\sqrt{3}}(\delta \tilde{N}_{3}-\delta 
\tilde{N}_{2}).
\end{eqnarray}%
We replace $\delta \tilde{N}_{1,2,3}$ by $\delta \tilde{S}_{\pm }$ as the
variables for the linearised equations. Note that the relation (\ref%
{eq:deltaN}) was used to reduce the number of variables. Finally, the
evolution of the perturbations are given by 
\begin{eqnarray}
\delta \tilde{H}^{\prime } &=&\frac{3\gamma }{2}\delta \tilde{\Omega}%
_{\Lambda } \\
\delta \tilde{\Omega}_{\Lambda }^{\prime } &=&2\left( 1-\frac{2}{3\gamma }%
\right) \delta \tilde{H} \\
\delta \tilde{\Sigma}_{\pm }^{\prime } &=&-\delta \tilde{S}_{\pm } \\
\delta \tilde{S}_{\pm }^{\prime } &=&8\delta \tilde{\Sigma}_{\pm }.
\end{eqnarray}%
The linear stability of the static solution is therefore decided by the
eigenvalues: 
\begin{equation}
\lambda =\pm \sqrt{3\gamma -2},\ \ \ \ \ \pm i2\sqrt{2},\ \ \ \ \ \pm i2%
\sqrt{2}.
\end{equation}%
We see that there is always an instability in the direction of increasing
and decreasing $\tau $ time for $\gamma >2/3$ because of the two real
eigenvalues with opposite signs. Note that this mode is associated with the
volume expansion (it is the one identified in ref \cite{BEMT}), namely the
FLRW mode and would still be present in the isotropic limit where $\tilde{N}%
_{1,2,3}=\tilde{\Sigma}_{\pm }=0.$ The effects of the anisotropic curvature
are not decisive at this order and produce the two pairs of purely imaginary
eigenvalues. We note that two of the imaginary eigenvalues correspond to an
axisymmetric mode ($\delta \tilde{\Sigma}_{+}-\delta \tilde{S}_{+}$)
mentioned by \cite{uggla} who studied only the locally rotationally
symmetric (LRS) case, although their analysis considered occurrence of chaos 
\cite{barrow} following evolution away from the Einstein static universe.

\section{The Extension to Tilted Fluids}

We can add a tilt to the fluid velocity \cite{tilt}, so that the unit normal
of the homogeneous hypersurfaces $n^{a}$ and the fluid 4-velocity $u^{a}$
are not coincident, but differ by the addition of a peculiar velocity vector 
$v^{a}$, with associated scalar $v$, so that 
\[
u^{a}=\Gamma (n^{a}+v^{a})\ \ \ \ \ \Gamma =(1-v^{2})^{-\frac{1}{2}}.
\]%
Denote the energy density of the fluid seen by the observer $u^{a}$ by $\rho 
$ and it satisfies the equation of state (\ref{eq:eos}). If we decompose its
energy-momentum tensor with respect to $n^{a}$ as 
\[
T_{ab}=\mu n_{a}n_{b}+2q_{(a}n_{b)}+p(g_{ab}+n_{a}n_{b})+\pi _{ab}
\]%
such that $q_{a}n^{a}=\pi _{a}^{\ a}=0$ and $\pi _{ab}n^{b}=0$, then these
quantities are given by 
\begin{eqnarray}
\mu  &=&\left( 1+\gamma \Gamma ^{2}v^{2}\right) \rho ,  \label{eq:mu} \\
p &=&\left( \gamma -1+\frac{\gamma }{3}\Gamma ^{2}v^{2}\right) \rho , \\
q_{a} &=&\gamma \Gamma ^{2}\rho v_{a}, \\
\pi _{ab} &=&\gamma \Gamma ^{2}\left[ v_{a}v_{b}-\frac{1}{3}%
v^{2}(g_{ab}+n_{a}n_{b})\right] .  \label{eq:pi}
\end{eqnarray}%
In this case, the orthogonal equations in \cite{HUW} are no longer usable
and we start from the general equations for an arbitrary group invariant
orthonormal frame given in \cite{ESW}. Since the fluid velocity gives rise
to an energy flux with respect to the observer $n^{a}$, from the equations
(1.93) in \cite{ESW}, we cannot diagonalise $\sigma _{\alpha \beta }$ and $%
n_{\alpha \beta }$ simultaneously. Although we could use the remaining gauge
freedom to choose the spatial triad to diagonalise $n_{\alpha \beta }$, this
does not turn out to be the best gauge choice for analyzing the stability of
isotopic equilibrium points, as we will see. Instead, we proceed without
fixing the gauge, which means that we need to modify the definition of the
normalisation factor $D$. Noting that for an arbitrary matrix $A$, 
\begin{equation}
\frac{d}{dx}\det A=\det A\ \mathrm{tr}\left( A^{-1}\frac{dA}{dx}\right) ,
\label{eq:matrix}
\end{equation}%
it is easy to see from the equation (1.96) in \cite{ESW} that 
\begin{eqnarray*}
\frac{d}{dt}(\det n_{\alpha \beta }) &=&\det n_{\alpha \beta }\
(n^{-1})^{\mu \nu }\frac{dn_{\mu \nu }}{dt} \\
&=&-3H\det n_{\alpha \beta }
\end{eqnarray*}%
and so if we redefine $\hat{D}$ as 
\[
\hat{D}\equiv \sqrt{H^{2}+\frac{1}{4}\left( \det n_{\alpha \beta }\right) ^{%
\frac{2}{3}}},
\]%
then the normalisation will work as before. Let us first define the
normalised geometrical quantities by 
\[
\hat{H}\equiv \frac{H}{\hat{D}},\ \ \ \ \ \hat{\Sigma}_{\alpha \beta }\equiv 
\frac{\sigma _{\alpha \beta }}{\hat{D}}\ \ \ \ \ \hat{N}_{\alpha \beta
}\equiv \frac{n_{\alpha \beta }}{\hat{D}}.
\]%
There now exists an angular velocity, $\Omega _{\alpha }$, of the frame with
respect to the Fermi-propagated frame. This is non-dynamical and is just a
manifestation of the gauge freedom. It is normalised by 
\[
\hat{R}_{\alpha }\equiv \frac{\Omega _{\alpha }}{\hat{D}}.
\]%
The fluid density parameter seen by the observer $n^{a}$ is defined by 
\[
\hat{\Omega}_{M}\equiv \frac{\mu }{3\hat{D}^{2}}=\left( 1+\gamma \Gamma
^{2}v^{2}\right) \frac{\rho }{3\hat{D}^{2}}.
\]%
Since the tilt velocity, $v_{\alpha }$, is already dimensionless, it will
not be normalised. The cosmological constant represents another dynamical
degree of freedom through the variable 
\[
\hat{\Omega}_{\Lambda }\equiv \frac{\Lambda }{3\hat{D}^{2}}.
\]%
We can now derive a new system of equations from the equations
(1.90)-(1.100) in \cite{ESW}.

First of all, the evolution of $\hat{D}$ decouples as before; 
\begin{equation}
\hat{D}^{\prime }=-(1+\hat{q})\hat{H}^{2}\hat{D}
\end{equation}%
where $^{\prime }$ now denotes $d/d\hat{\tau}=\hat{D}^{-1}d/dt$ and $\hat{q}$
is determined by 
\begin{equation}
\hat{q}=2\hat{\Sigma}^{\mu \nu }\hat{\Sigma}_{\mu \nu }+\frac{3\gamma
-2-(\gamma -2)v^{2}}{2(1+(\gamma -1)v^{2})}\hat{\Omega}_{M}-\hat{\Omega}%
_{\Lambda }.
\end{equation}%
The rest of the Einstein equations read 
\begin{eqnarray}
\hat{\Sigma}_{\alpha \beta }^{\prime } &=&-(2-\hat{q})\tilde{H}\hat{\Sigma}%
_{\alpha \beta }+2\epsilon _{\ \ (\alpha }^{\mu \nu }\hat{\Sigma}_{\beta
)\mu }\hat{R}_{\nu }-\hat{S}_{\alpha \beta }+\hat{\Pi}_{\alpha \beta }, \\
1 &=&\hat{\Sigma}^{\mu \nu }\hat{\Sigma}_{\mu \nu }+\hat{V}+\hat{\Omega}_{M}+%
\hat{\Omega}_{\Lambda },  \label{eq:const3} \\
0 &=&\frac{3\gamma \hat{\Omega}_{M}}{1-(\gamma -1)v^{2}}v_{\alpha }+\epsilon
_{\alpha }^{\ \mu \nu }\hat{\Sigma}_{\mu }^{\ \beta }\hat{N}_{\beta \nu },
\label{eq:const4}
\end{eqnarray}%
where $\epsilon _{\mu \nu \sigma }$ is the spatial Levi-Civita symbol and 
\begin{eqnarray}
\hat{S}_{\alpha \beta } &=&2\hat{N}_{\alpha }^{\ \mu }\hat{N}_{\mu \beta }-%
\frac{2}{3}\hat{N}_{\mu \nu }\hat{N}^{\mu \nu }\delta _{\alpha \beta }-\hat{N%
}_{\ \mu }^{\mu }\left[ \hat{N}_{\alpha \beta }-\frac{1}{3}\hat{N}_{\ \mu
}^{\mu }\delta _{\alpha \beta }\right] , \\
\hat{\Pi}_{\alpha \beta } &=&\frac{3\gamma \hat{\Omega}_{M}}{1-(\gamma
-1)v^{2}}\left[ v_{\alpha }v_{\beta }-\frac{1}{3}v^{2}\delta _{\alpha \beta }%
\right]  \\
\hat{V} &=&\frac{1}{4}\left( \det \hat{N}_{\alpha \beta }\right) ^{\frac{2}{3%
}}+\frac{1}{6}\hat{N}_{\mu \nu }\hat{N}^{\mu \nu }-\frac{1}{12}\left( \hat{N}%
_{\ \mu }^{\mu }\right) .
\end{eqnarray}%
The Jacobi identities become 
\begin{equation}
\hat{N}_{\alpha \beta }^{\prime }=\hat{q}\tilde{H}\hat{N}_{\alpha \beta }+2%
\hat{\Sigma}_{(\alpha }^{\ \ \mu }\hat{N}_{\beta )\mu }+2\epsilon _{\ \
(\alpha }^{\mu \nu }\hat{N}_{\beta )\mu }\hat{R}_{\nu }.
\end{equation}%
The evolution for the matter variables is obtained by substituting (\ref%
{eq:mu})-(\ref{eq:pi}) into the contracted Bianchi identities: 
\begin{eqnarray}
\hat{\Omega}_{M}^{\prime } &=&2(1+\hat{q})\hat{H}\hat{\Omega}_{M}-\frac{%
\gamma }{1+(\gamma -1)v^{2}}\left[ (3+v^{2})\hat{H}+\hat{\Sigma}_{\alpha
\beta }v^{\alpha }v^{\beta }\right] \hat{\Omega}_{M}, \\
v_{\alpha }^{\prime } &=&\frac{\Gamma ^{-2}(3\gamma -4)\tilde{H}-(\gamma -2)%
\hat{\Sigma}_{\mu \nu }v^{\mu }v^{\nu }}{1-(\gamma -1)v^{2}}v_{\alpha } \\
&&-\hat{\Sigma}_{\alpha \beta }v^{\beta }-\epsilon _{\alpha \mu \nu }\hat{R}%
^{\mu }v^{\nu }+\epsilon _{\alpha \mu \nu }\hat{N}_{\ \beta }^{\mu }v^{\beta
}v^{\nu },  \nonumber \\
\hat{\Omega}_{\Lambda }^{\prime } &=&2(1+\hat{q})\tilde{H}\hat{\Omega}%
_{\Lambda }
\end{eqnarray}%
Finally, by the definition of the normalisation, we have 
\begin{eqnarray}
\tilde{H}^{\prime } &=&-(1-\tilde{H}^{2})\hat{q}, \\
1 &=&\tilde{H}^{2}+\frac{1}{4}\left( \det \hat{N}_{\alpha \beta }\right) ^{%
\frac{2}{3}},  \label{eq:const5}
\end{eqnarray}

The normalised fluid density, $\hat{\Omega}_{M}$, is usually eliminated by
the constraint (\ref{eq:const3}) and then there are 16 dynamical variables
(taking into account the traceless condition for $\hat{\Sigma}_{\alpha \beta
}$). We still have 3 degrees of gauge freedom and 4 constraints ((\ref%
{eq:const4}) and (\ref{eq:const5})), so the dynamical system is
9-dimensional, which is consistent with the counting of the ref \cite{siklos}
because now we have one more degree of freedom coming from addition of the
cosmological constant to the 8 in general single-fluid tilted models. \emph{%
\ }

To find equilibrium points in the system, we need consider the gauge. Unless
we exhaust this freedom, an equilibrium point can be time dependent. A
typical choice is to diagonalize $\hat{N}_{\alpha \beta }$. It leads to the
following defining equations for the non-dynamical $\hat{R}_{\alpha }$: 
\begin{eqnarray*}
\hat{R}_{1} &=&\frac{\hat{\Sigma}_{23}(\hat{N}_{22}+\hat{N}_{33})}{\hat{N}%
_{22}-\hat{N}_{33}} \\
\hat{R}_{2} &=&\frac{\hat{\Sigma}_{13}(\hat{N}_{33}+\hat{N}_{11})}{\hat{N}%
_{33}-\hat{N}_{11}} \\
\hat{R}_{3} &=&\frac{\hat{\Sigma}_{12}(\hat{N}_{11}+\hat{N}_{22})}{\hat{N}%
_{11}-\hat{N}_{22}}.
\end{eqnarray*}%
If we substitute those expressions into the evolution equations, they become
singular when the denominators vanish, which includes the isotropic case.
Thus this gauge is not suited for the stability analysis of Einstein static
universe. The situation is the same for a gauge where $\hat{\Sigma}_{\alpha
\beta }$ is diagonal.

Fortunately, in the Einstein static universe all the dynamical variables
have to be time independent since all the gauge dependent variables are
zero. While there is ambiguity in the values of $\hat{R}_{\alpha }$ which
are arbitrary (potentially time-dependent) since they merely represent the
rotation of the spatial frame aside from it, the equilibrium point is gauge
invariantly characterised by 
\[
\hat{H}=0,\ \ \ \ \ \hat{\Sigma}_{\alpha \beta }=0,\ \ \ \ \ \hat{N}_{\alpha
\beta }=2\delta _{\alpha \beta },\ \ \ \ \ \ \hat{\Omega}_{M}=\frac{2}{%
3\gamma }=1-\hat{\Omega}_{\Lambda },\ \ \ \ \ v_{\alpha }=0.
\]%
Let us consider a perturbation around this background without fixing the
gauge. In this way, we expect to have 12 eigenvalues taking into account the
four constraints mentioned above. From now on, the quantities without $%
\delta $ prefixes are understood to take the background values. The
constraint (\ref{eq:const4}) becomes 
\[
2\delta v_{\alpha }=-\epsilon _{\alpha }^{\ \mu \nu }\delta \hat{\Sigma}%
_{\mu }^{\ \beta }\hat{N}_{\beta \nu }\equiv 0.
\]%
Thus, the velocity perturbation to linear order automatically vanishes, and
so $\delta \hat{\Pi}_{\alpha \beta }=0$. Secondly, the constraint (\ref%
{eq:const5}) implies 
\begin{eqnarray*}
0 &=&\delta (\det \hat{N}_{\alpha \beta }) \\
&=&\frac{\partial \det \hat{N}_{\alpha \beta }}{\partial \hat{N}_{\mu \nu }}%
\delta \hat{N}_{\mu \nu } \\
&=&\det \hat{N}_{\alpha \beta }\mathrm{tr}\left( (\hat{N}^{-1})_{\alpha }^{\
\rho }\frac{\partial \hat{N}_{\rho \beta }}{\partial \hat{N}_{\mu \nu }}%
\right) \delta \hat{N}_{\mu \nu }
\end{eqnarray*}%
where we used (\ref{eq:matrix}) again. Noting that 
\[
(\hat{N}^{-1})_{\alpha \beta }=\frac{1}{2}\delta _{\alpha \beta }
\]%
and 
\[
\frac{\partial \hat{N}_{\rho \sigma }}{\partial \hat{N}_{\mu \nu }}=\delta
_{\ \rho }^{\mu }\delta _{\ \sigma }^{\nu },
\]%
we can see that 
\[
\delta \hat{N}_{\mu }^{\ \mu }=0,
\]%
namely, $\delta \hat{N}_{\alpha \beta }$ is trace-free. This ensures that $%
\delta \hat{V}=0$. It's also easy to see that $\delta \hat{S}_{\alpha \beta
}=2\delta \hat{N}_{\alpha \beta }$. Therefore, the linearised evolution
equations are quite simple: 
\begin{eqnarray*}
\delta \hat{\Sigma}_{\alpha \beta }^{\prime } &=&-2\delta \hat{N}_{\alpha
\beta }+2\epsilon _{\ \ (\alpha }^{\mu \nu }\delta \hat{\Sigma}_{\beta )\mu }%
\hat{R}_{\nu }, \\
\delta \hat{N}_{\alpha \beta }^{\prime } &=&4\delta \hat{\Sigma}_{\alpha
\beta }+2\epsilon _{\ \ (\alpha }^{\mu \nu }\delta \hat{N}_{\beta )\mu }\hat{%
R}_{\nu }, \\
\delta \hat{H}^{\prime } &=&\frac{3\gamma }{2}\delta \hat{\Omega}_{\Lambda },
\\
\delta \hat{\Omega}_{\Lambda }^{\prime } &=&2\left( 1-\frac{2}{3\gamma }%
\right) \delta \hat{H}.
\end{eqnarray*}%
The equations decouple into two. The scalar part is exactly the same as the
non-tilted case and gives one positive and one negative eigenvalue and so
the Einstein static universe is also unstable under the same conditions as
in the non-tilted case. Therefore there is instability of the static
solution for increasing and decreasing times. The tensorial part can be
brought into a standard form 
\begin{eqnarray*}
\delta \hat{\Sigma}_{\alpha \beta }^{\prime } &=&-2\delta \hat{N}_{\alpha
\beta } \\
\delta \hat{N}_{\alpha \beta }^{\prime } &=&4\delta \hat{\Sigma}_{\alpha
\beta }
\end{eqnarray*}%
by an appropriate rotation of the spatial frame, leaving five pairs of
coupled equations, and therefore five sets of eigenvalues $\pm i2\sqrt{2}$.
Although not all of them are physical as we haven't fix the gauge, this is
unimportant because all the eigenvalues are purely imaginary. This means the
nature of the tensorial perturbations is also the same as in the non-tilted
case.

\section{Conclusions}

In this study of the stability of the Einstein static universe against
Bianchi type IX modes we have extended the earlier studies of ref \cite{BEMT}%
, that considered only the destabilizing effects of a single homogeneous
mode in the presence of comoving fluids, and those of ref \cite{uggla}, who
considered only the evolution of LRS type IX models with non-tilted fluid.
We investigated the general type IX evolution in the vicinity of the
Einstein static model in the presence of a fluid with non-tilted and tilted
fluid motion. We have established that the Einstein static universe is
unstable to Bianchi type-IX spatially homogeneous perturbations in the
presence of non-tilted and tilted perfect fluids with $\rho +3p>0$. We also
found that the imaginary eigenvalues corresponding to the perturbative
effects of anisotropic curvature (Mixmaster modes) and fluid tilt generalise
the oscillatory behaviour of the finite wavelength vector and tensor
perturbations found in early studies of small amplitude perturbations.

\end{document}